# Early Detection of Parkinson's Disease through Patient Questionnaire and Predictive Modelling

R. Prashanth[a,*], Sumantra Dutta Roy[a]

[a] Department of Electrical Engineering, Indian Institute of Technology Delhi, India

**Abstract**

Early detection of Parkinson's disease (PD) is important which can enable early initiation of therapeutic interventions and management strategies. However, methods for early detection still remain an unmet clinical need in PD. In this study, we use the Patient Questionnaire (PQ) portion from the widely used Movement Disorder Society-Unified Parkinson's Disease Rating Scale (MDS-UPDRS) to develop prediction models that can classify early PD from healthy normal using machine learning techniques that are becoming popular in biomedicine: logistic regression, random forests, boosted trees and support vector machine (SVM). We carried out both subject-wise and record-wise validation for evaluating the machine learning techniques. We observe that these techniques perform with high accuracy and high area under the ROC curve (both >95%) in classifying early PD and healthy normal. The logistic model demonstrated statistically significant fit to the data indicating its usefulness as a predictive model. It is inferred that these prediction models have the potential to aid clinicians in the diagnostic process by joining the items of a questionnaire through machine learning.

[*]Corresponding author Email: prashanth.r.iitd@gmail.com

Abbreviations: PD− Parkinson's disease, PPMI− Parkinson's Progression Markers Initiative, PQ− Patient Questionnaire, SVM− Support Vector Machine, MDS-UPDRS− Movement Disorder Society-Unified Parkinson's Disease Rating Scale

# 1. Introduction

Parkinson's disease (PD) is a neurodegenerative disorder affecting millions of elderly people, significantly affecting their quality of lives [1]. PD is a complex disorder characterized by several motor and non-motor symptoms that worsen over time. In advanced stages of PD, clinical diagnosis is clear-cut. However, in the early stages, when the symptoms are often incomplete or subtle, the diagnosis becomes difficult and at times, the subject may remain undiagnosed. For instance, a large-scale epidemiological study on a European cohort showed that a significant proportion of previously undetected subjects, were diagnosed with PD, after screening them with a symptom questionnaire followed by physical examination and clinical intervention [2]. Tremor, the most common symptom in PD, may never manifest in some patients and the patient may show other relevant symptoms such as bradykinesia (slower and smaller handwriting, decreased arm swing and leg stride when walking, decreased facial expression, and decreased amplitude of voice) to begin with [1]. The difficulty in early detection of PD is a strong motivation for computer-based assessment tools/decision support tools/test instruments that can aid in the early diagnosis of PD. Timely detection, preferably at a stage earlier than currently possible, and subsequent intervention could be hugely beneficial in a way that the patient could have access to disease modifying therapy to slow down the course of PD progression.

Machine learning provides seemingly immense opportunities to computer-aided classification and diagnosis that could reduce inevitable fallibilities and inherent diagnostic variabilities in healthcare, provide guidance (especially in a setting when expert physicians are not available), and speed up decision making. These models can aid in the early detection of PD and also for identifying subjects for clinical trials. Researchers have used a variety of data for solving the PD detection problem via machine learning techniques. [3-7] used speech data, [8] studied gait patterns, [9] carried out analysis on force tracking data, [10-12] performed analysis on single photon emission computed tomography (SPECT) scan data, and [13] used smell identification data to distinguish PD from healthy normal. [14] carried out assessment of PD state based on many weeks of movement data collected from sensors from 34 subjects who were in mild to severe PD. [15] describes about the collaboration between Michael J. Fox Foundation for PD Research and Intel Corporation, to develop a mobile application and an Internet of Things (IoT) platform to support large-scale studies of objective, continuously sampled sensory data from people with PD. [16] in their study, proposes an approach to evaluate the feasibility and compliance of using multiple wearable sensors to collect clinically relevant data, and to address the usability of these data for answering clinical research questions. [17] is a continuation of their previous study [16] where they complete their data collection process and show that it is feasible to deploy a technology platform consisting of consumer-grade wearable and mobile devices to capture large amounts of sensor data from many participants from a large and geographically diverse PD population. [18] presents an approach to analyze smartwatch data from 19 PD subjects and shows high performance in terms of detecting symptoms of tremor, bradykinesia, and dyskinesia. [19] showed that dual-wrist sensor fusion may enable robust gait quantification (aiding in capturing timing-based, gait abnormalities) in free-living environments. [20] provides a review which discusses promising wearable technology, addresses which parameters should be prioritized in such assessment strategies, and reports about studies that have already investigated daily life issues in PD using this new technology. [21] used imaging, genetics, clinical and demographic data to develop prediction models for PD. [22] developed predictions models for PD using genetic, non-motor and demographic data. [23] used motor, non-motor and imaging data from to develop prediction models for PD. [24] developed multi-task learning model for the prediction of PD progression, measured using baseline measurements of biologic specimen, clinical assessments and brain imaging. [25] explored the non-motor symptoms (NMS) and quality of life (QOL) in tremor dominant (TD) vs. postural instability gait disorder (PIGD) PD patients, and observed that PIGD phenotype had a higher prevalence of NMS and worse QOL than TD phenotype. These studies had one or more of the following shortcomings: they used only the partial aspects of PD (for example, using only motor aspects like speech or movement or non-motor aspects like the difficulty in olfaction) and did not make use of broad spectrum of symptoms in PD (both motor and non-motor); their sample size was limited; the accuracy of detecting PD was low; it required special hardware, sensors and sophisticated techniques for acquiring the data like the speech and movement data which makes their implementation costly and difficult; and/or they used expensive methods involving SPECT scanning. In our previous work [26], we used a combination of non-motor features of olfactory loss and Rapid Eye Movement sleep Behavior Disorder, along with other relevant biomarkers of Cerebrospinal fluid (CSF) measurements and SPECT imaging markers to develop predictive models for detecting early PD.

Previous studies based on Patient Questionnaires (PQ) for PD have shown that the questionnaire features capture significant amount of information that helps in diagnosis [27-30]. The Movement Disorder Society-Unified Parkinson's Disease Rating Scale (MDS-UPDRS) [31] is a widely used rating scale evaluating both motor as well as non-motor aspects of PD, and estimating its severity and progression. MDS-UPDRS consists of 4 parts and among which a portion of Part I (Parts IB, hereafter) and Part II, containing a total of 20 questions, is in the form of a patient questionnaire. Part IB assesses the non-motor aspects of daily living and Part II assess the motor aspects of daily living. Studies [31-34] have validated these parts and have observed that they provide a relevant estimate of PD symptoms. The Hoehn and Yahr (HY) scale is another widely used scale for assessing PD stage [35]. This scale provides an overall assessment of severity through staging. The progression of PD usually starts from unilateral (Stage 1), to bilateral without balance difficulties (Stage 2), followed by the presence of postural instability (Stage 3), to loss of physical independence (Stage 4), and being wheel-chair or bed-bound unless aided (Stage 5). The HY scale has also been used to categorize PD as early stage (Stages 1 and 2), moderate stage (Stage 3) and late stage (Stages 4 and 5). Studies have shown that the PD stage has a significant correlation with quality of life measures and the UPDRS [36]. We have used the data from healthy normals and PD subjects who are in their early stages in our analysis. Developing predictive models that can perform classification or compute the likelihood of PD using patient questionnaire features is the approach that we carry out in this paper. There have been attempts to relate UPDRS (previous version of MDS-UPDRS) and the severity of PD through estimating the HY stage [37-39]. Scanlon *et al.* [37, 38] propose formulas to obtain HY stages from UPDRS Part III scores. Tsanas *et al.* [39] optimize this formula by refining its parameters using genetic algorithm (GA). However, these studies do not carry out predictive modelling and their formula was based on intuitive rules, and furthermore, Part III of MDS-UPDRS which they used requires an expert PD clinician for rating.

There are multiple advantages of using PQ features from MDS-UPDRS: a) it is simple to understand [31], they are easy to administer and can be administered even by primary physicians who are not PD experts , it is cheap and does not involve any invasive procedures; b) they have been extensively tested and validated in patient groups, and are observed to reflect an effective and relevant estimate of broad spectrum of symptoms in PD [31-34]; c) subjects can fill the questionnaire themselves. Studies based on PQ parts of UPDRS [40] (MDS-UPDRS is the updated version of UPDRS) show that PD patient self-assessment and caregiver evaluation of the patient's disability showed close concordance with the neurologist's ratings, and that they are a reliable and valid outcome measure [41, 42]; d) it assesses both non-motor (Part-IB) and motor (Part-II) features of PD [31].

In this paper, we use the PQ parts (Parts IB and II) of MDS-UPDRS [31]) along with demographic information to classify PD patients from healthy normal using logistic regression, random forests, boosted trees and support vector machine (SVM).

## 2. Materials and Methods

### 2.1. Dataset details

The data used for the study was from the Parkinson's Progression Markers Initiative (PPMI) database (www.ppmi-info.org/data). For up-to-date information on the study, please visit www.ppmi-info.org. PPMI [43] is a landmark, large-scale, observational and multi-center study that recruits early-untreated PD patients along with age- and gender-matched healthy normal subjects, to identify progression biomarkers in PD.

We have used the PQ portion of the MDS-UPDRS for the analysis. The complete MDS-UPDRS is freely available online at https://www.movementdisorders.org/MDS-Files1/PDFs/Rating-Scales/MDS-UPDRS_English_FINAL.pdf. And the research article [31] providing more details on MDS-UPDRS can be found at http://onlinelibrary.wiley.com/doi/10.1002/mds.22340/pdf.

The PQ parts of the MDS-UPDRS consisting of a total of 7 (in Part IB) + 13 (in Part II) = 20 evaluations along with age and gender information, making a total of 22 features, are used as the feature set in this study. Each evaluation is answered based on its severity (0 – normal; 1 – slight; 2 – mild; 3 – moderate; 4 – severe). The evaluations in Part IB are the following:

*1. Sleep Problems, 2. Daytime Sleepiness, 3. Pain and other sensations, 4. Urinary Problems, 5. Constipation problems, 6. Light Headedness on standing, 7. Fatigue.*

The evaluations in Part II are the following:

*8. Speech, 9. Saliva and Drooling, 10. Chewing and Swallowing, 11. Eating tasks, 12. Dressing, 13. Hygiene, 14. Handwriting, 15. Doing Hobbies and other activities, 16. Turning in bed, 17. Tremor, 18. Getting out of bed/car/deep chair, 19. Walking and balance, 20. Freezing.*

PPMI is a longitudinal study where both PD and healthy subjects undergo a comprehensive longitudinal schedule of assessments [43]. Evaluations occur at screening/baseline and at 3 month intervals during the first year of participation and then every 6 months thereafter for the next four years and it is followed by yearly assessment thereafter. The dataset contained observations for close to 7 years (July 2010 to April 2017). Average number of follow up visits for healthy normal is 5.06 and for PD is 9.92. The dataset consisted of a total of 5704 observations with 1002 samples from 198 healthy normal subjects, and 4702 samples from 474 early PD [44] (Hoehn and Yahr stage 1 or 2 with mean as 1.62) patients. The PQ was filled either by the patient / caregiver / patient and caregiver combined. Table 1 shows the distribution for each category. It is observed that around 99% of the healthy normal subjects and 95% early PD subjects put the ratings themselves.

**Table 1: Distribution of how the PQ was filled**

|  | N(%) | |
|---|---|---|
|  | *Healthy Normal* | *Early PD* |
| *Patient* | 994 (99.20%) | 4475 (95.17%) |
| *Caregiver* | 6 (0.60%) | 8 (0.17%) |
| *Patient and Caregiver combined* | 2 (0.20%) | 219 (4.66%) |
| *Total* | 1002 | 4702 |

Figure 1(a) shows the stacked bar charts depicting the severity of features for normal and Figure 1(b) for early PD groups. It can be observed that all the features are getting affected in PD. Figure 2 shows the quantitative measure for the affected features. The quantification is done by computing the difference of percentages of samples showing normal behavior for each feature between Healthy Normal and Early PD. From this plot, we can observe that tremor, handwriting and dressing are the most affected features. As the PPMI study mainly recruits early and untreated PD subjects, their symptoms or the features (in the questionnaire) are subtle. For instance, the number of early PD observations showing no sign of tremor (which is the most obvious symptom in PD, feature no. 17) is 715 which is around 15% of total early PD observations in the study. Figure 3 shows the percentage of PD samples showing normal behavior for each feature. From this plot, it is clear that majority of the observations from the PD group showed severity in tremor, and only few observations showed severity in the Freezing symptom.

**2.2. Data Partitioning**

As PPMI is a longitudinal study where subjects are followed up, measurements are taken at different time points for each subject. In this study, along with record-wise cross-validation, we also carry out subject-wise cross-validation. In subject-wise cross-validation, the data is partitioned in a 10-fold cross-validation set up such that for each fold about 90% subjects, i.e. data from 605 subjects (178 Normal + 427 Early PD) were used for training and rest about 10% subjects, i.e., data from 67 subjects (20 Normal + 47 Early PD) were used for testing, and this cross-validation process is repeated 100 times. In record-wise cross-validation, the data is partitioned in a 10-fold cross-validation set up such that for each fold 90% of the whole data, i.e., 5134 observations were used for training and rest 10%, i.e., 570 observations were used for testing, and this cross-validation process is repeated 100 times The partitioning of the dataset into training and testing sets is carried out in a stratified manner, i.e., in such a way that both sets are mutually exclusive and have roughly the same class proportions as in the set of class labels. The average measure from the 100 cross-validation runs is taken to get an unbiased estimate of performance measures used which are accuracy, sensitivity, specificity and area under the ROC curve (AUC).

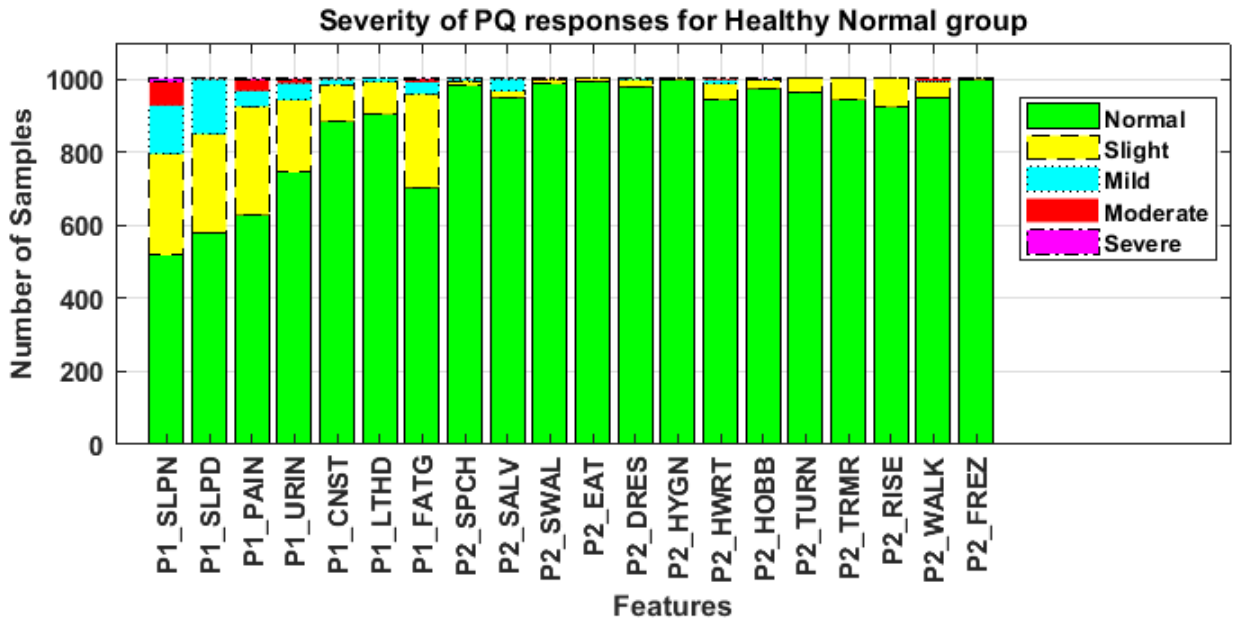

(a)

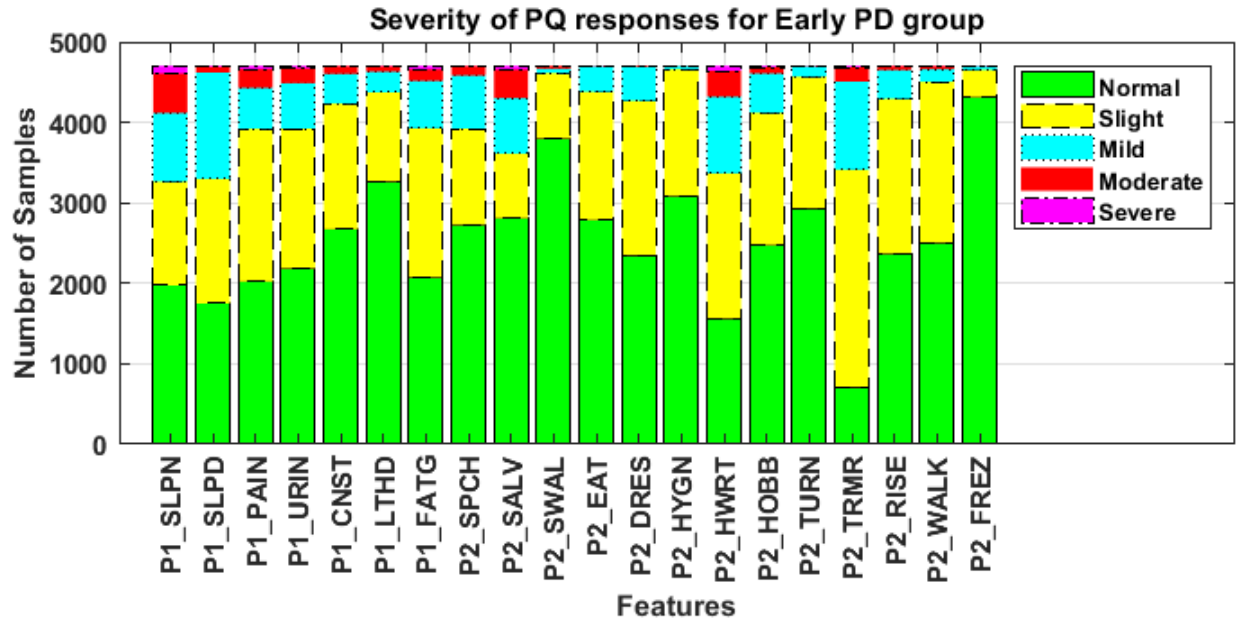

(b)

Fig. 1. Stacked bar charts showing the severity of PQ features in the MDS-UPDRS for (a) Healthy Normal, and (b) Early PD groups. $X$- and $Y$-axes represent the features (abbreviated) and the number of observations, respectively. The features from Part I are indicated by P1 and those from Part II are indicated by P2. The severity is indicated by color-coding as indicated by the legend in the figure. The plots show that all features are getting affected in PD.

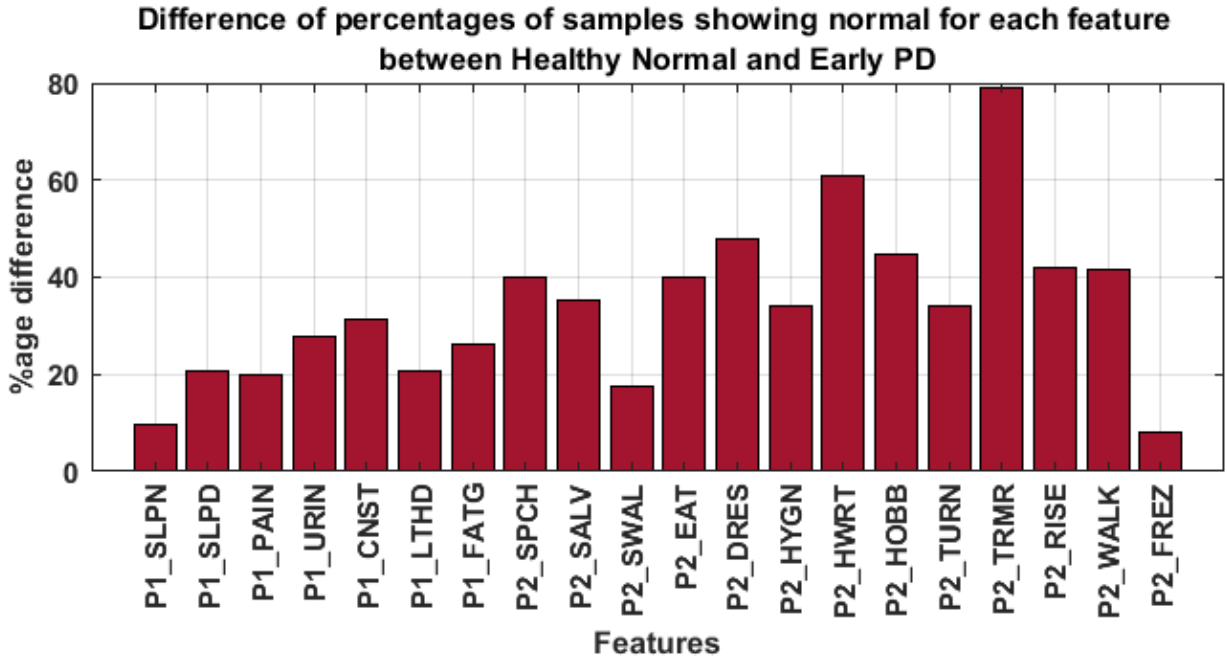

Fig. 2. Plot of difference of percentages of observations showing normal behavior for each feature between the Normal and PD groups. It is observed that tremor (P2_TRMR), handwriting (P2_HWRT) and dressing (P2_DRESS) were the most affected features.

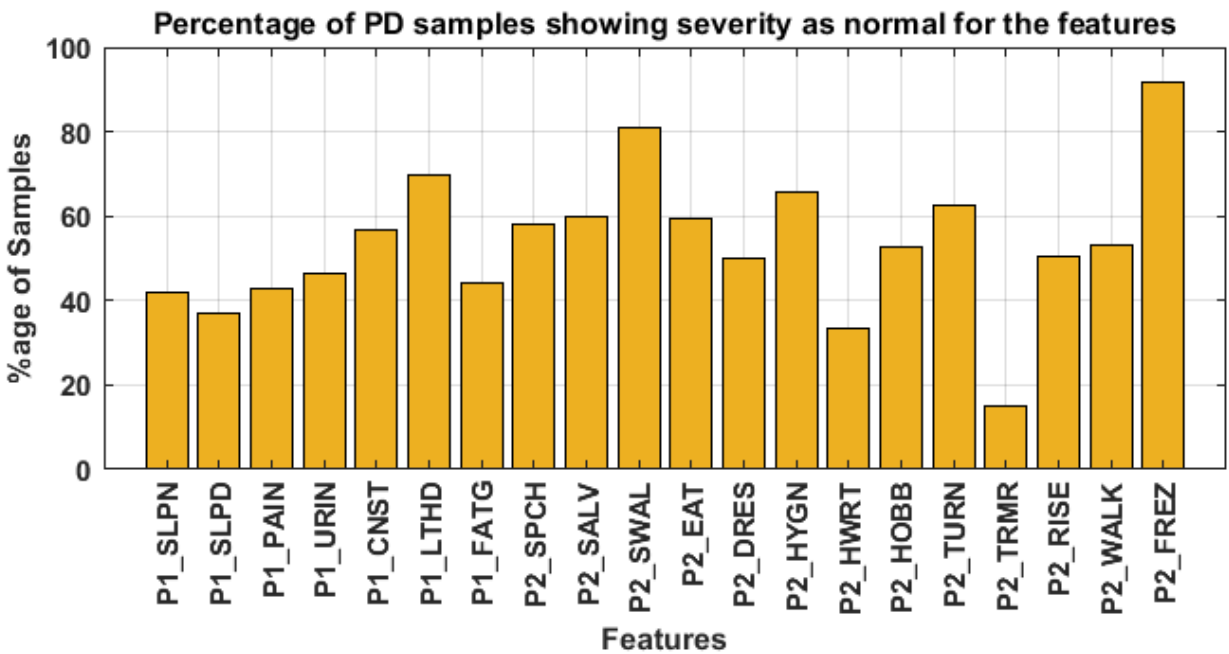

Fig. 3. Plot of percentage of samples from PD group who show normal behavior for each feature.

**2.3 Feature selection techniques used**

Feature selection is a necessary step before predictive modelling when there is redundancy between the features. It is also necessary as decreased set of features reduces the size of the problem which makes the training time lesser, and there is lesser chance of overfitting. To carry out feature selection, we use the techniques as mentioned below:

**2.3.1 Wilcoxon rank sum test**

It is a filter based method where the selection of features is carried out based on (univariate) statistical tests. In this study, we use the Wilcoxon rank sum test for analyzing the significance of features. The Wilcoxon rank sum test is a nonparametric test for two populations to test for the equality of the population medians [45]. We use this test, instead of the *t-test* as most of the features do not show a normal distribution. For illustration, the histograms of the first four features are as shown below in Figure 4.

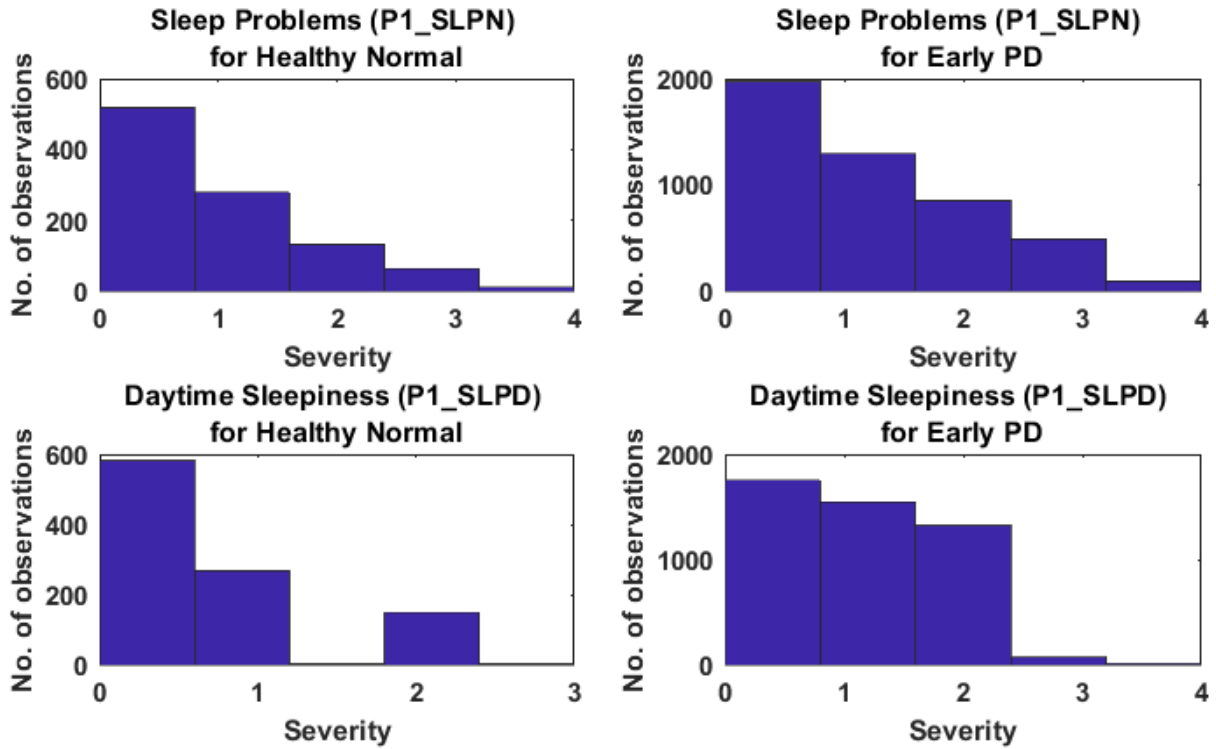

Fig. 4. Histograms of the first four features. We can clearly observe that the histograms are highly skewed. This is because the PD subjects in the study were all in their early stages where the symptoms were less severe.

### 2.3.2 LASSO based feature selection

Least Absolute Shrinkage and Selection Operator (LASSO) is a regularization technique which applies a shrinking process where it penalizes the coefficients of the regression variables shrinking some of them to zero [46]. During this process, the variables that still have a non-zero coefficient are selected for subsequent prediction modelling.

### 2.3.3 PCA

PCA is a dimensionality reduction technique used to decompose a multivariate dataset to a set of orthogonal components. It converts the dataset of possibly correlated variables into a set of values of linearly uncorrelated variables called principal components. These components are obtained from the eigenvectors of the covariance matrix, and give directions in which the data have maximal variance [47]. In that way, the first principal component has the largest possible variance. Features are selected such that the total percentage of variance explained by selected components is not less than 99%.

**2.4. Predictive modelling for distinguishing early PD from healthy normal**

The input data can be represented as $x_i = [x_{i_1}, x_{i_2}, \dots, x_{i_k}]$; $i = 1, \dots, n$; $k$ is number of features, $k = 22$, and $n$ is number of observations, $n=5704$; and the output class as $y \in \{0, 1\}$ where 0 and 1 represent normal and early PD, respectively. The goal of the predictive model is to compute $p(y|x_i)$ which is the output class (or probability of PD) for an observation $x_i$.

For classification, we use conventional techniques such as logistic regression [47] which is among the most popular techniques used in biomedical, as well as more recent techniques such as random forests [48, 49], boosted trees [50] and SVM [51] that are becoming very popular in biomedicine [49, 52]. We use statistics toolbox in MATLAB to carry out classification using logistic regression, random forests and boosted trees, and LIBSVM [53] library for SVM modelling. More detailed description of these techniques is given in the Supplementary file.

**2.5. Variable importance**

Along with classification, Random forests technique was also used to evaluate feature importance [48]. During bagging (which is selecting samples with replacement) process as used in Random forests, it leaves out about 37% of the examples for each tree. It means that each tree is trained using only about 63% of the data on average. These left out examples are called out-of bag (OOB) samples. The first step in computing the variable importance is to fit a Random forests model to the data. Compute the OOB error which is the misclassification rate for OOB observations in the data. After this, to estimate the importance of the $j^{th}$ feature, the values of the $j^{th}$ feature are randomly permuted (shuffled) in the training data and the OOB error is again computed on this perturbed data set. The importance score for the feature is computed by averaging the difference in OOB error before and after the permutation. The score is normalized by the standard deviation of these differences. Features with higher scores are ranked higher.

## 3. Results

### 3.1. Feature Selection

For the analysis, a 10-fold nested cross-validation based approach was carried out where the feature selection and parameter tuning happens in the inner fold and evaluation happens in an independent set (or fold).

#### 3.1.1 Features Selected from Wilcoxon rank sum test

The features are statistically tested using two-sided Wilcoxon rank sum test and 21 of them were observed to be statistically significant in all folds ($p$-value $\ll$ 0.05). Age was statistically insignificant as both the groups (Healthy normal and Early PD) were age matched. Table 2 below shows the results of this statistical testing. All features, except age, were included in the subsequent prediction modelling.

**Table 2: Statistical testing of features**

| SNo | Features | Abbrv. | HC (mean±SD) | PD (mean±SD) | z-stat | p-val |
|---|---|---|---|---|---|---|
| 1 | Sleep Problems | P1_SLPN | 0.77±0.97 | 1.03±1.09 | -6.7863 | 1.15E-11 |
| 2 | Daytime Sleepiness | P1_SLPD | 0.57±0.75 | 0.95±0.86 | -12.758 | 2.81E-37 |
| 3 | Pain and other sensations | P1_PAIN | 0.49±0.75 | 0.8±0.88 | -11.681 | 1.60E-31 |
| 4 | Urinary problems | P1_URIN | 0.33±0.63 | 0.75±0.86 | -15.911 | 5.35E-57 |
| 5 | Constipation problems | P1_CNST | 0.13±0.4 | 0.55±0.73 | -18.602 | 3.10E-77 |
| 6 | Light Headedness on standing | P1_LTHD | 0.11±0.34 | 0.39±0.67 | -13.581 | 5.20E-42 |
| 7 | Fatigue | P1_FATG | 0.35±0.6 | 0.77±0.85 | -15.625 | 4.89E-55 |
| 8 | Speech | P2_SPCH | 0.03±0.21 | 0.61±0.82 | -23.605 | 3.45E-123 |
| 9 | Saliva and Drooling | P2_SALV | 0.09±0.39 | 0.73±1.03 | -20.839 | 1.91E-96 |
| 10 | Chewing and Swallowing | P2_SWAL | 0.02±0.16 | 0.22±0.49 | -13.81 | 2.22E-43 |
| 11 | Eating tasks | P2_EAT | 0.01±0.08 | 0.48±0.63 | -24.15 | 7.54E-129 |
| 12 | Dressing | P2_DRES | 0.02±0.17 | 0.59±0.66 | -27.345 | 1.22E-164 |
| 13 | Hygiene | P2_HYGN | 0±0.05 | 0.35±0.5 | -21.677 | 3.40E-104 |
| 14 | Handwriting | P2_HWRT | 0.08±0.34 | 1.05±0.96 | -32.772 | 1.49E-235 |
| 15 | Doing Hobbies and other activities | P2_HOBB | 0.03±0.23 | 0.62±0.76 | -25.683 | 1.83E-145 |
| 16 | Turning in bed | P2_TURN | 0.04±0.19 | 0.41±0.55 | -20.965 | 1.37E-97 |
| 17 | Tremor | P2_TRMR | 0.06±0.23 | 1.17±0.74 | -43.366 | 0 |
| 18 | Getting out of bed/car/deep chair | P2_RISE | 0.08±0.27 | 0.59±0.68 | -24.216 | 1.50E-129 |
| 19 | Walking and balance | P2_WALK | 0.07±0.33 | 0.52±0.61 | -24.124 | 1.39E-128 |
| 20 | Freezing | P2_FREZ | 0±0.05 | 0.09±0.34 | -9.0586 | 1.32E-19 |
| 21 | Gender | GENDER | 0.62±0.49 | 0.66±0.47 | -2.3313 | 0.01974 |
| 22 | Age | AGE | 66.42±11.09 | 66.61±9.69 | 0.09147 | 0.927122 |

*p*-value shows statistical significance. Abbrv is the abbreviation.

### 3.1.2. Features selected from LASSO technique

A 10-fold cross-validated lasso regularization of a logistic regression model is created for feature selection. The features having non-zero coefficients were selected and the rest were discarded in each fold of the nested cross-validation.

### 3.1.3. Features selected from PCA

PCA decorrelates the features and finds a new representation from the existing features. Here, we have selected those many principal components which retain 99% of the total variance. Figure 5 as shown below visualizes the transformed representation using the first three features. The plot shows the largest variability along the first principal component axis. The second principal component axis has the second largest variability, which is significantly smaller than the variability along the first principal component axis. Similarly for the third principal component axis.

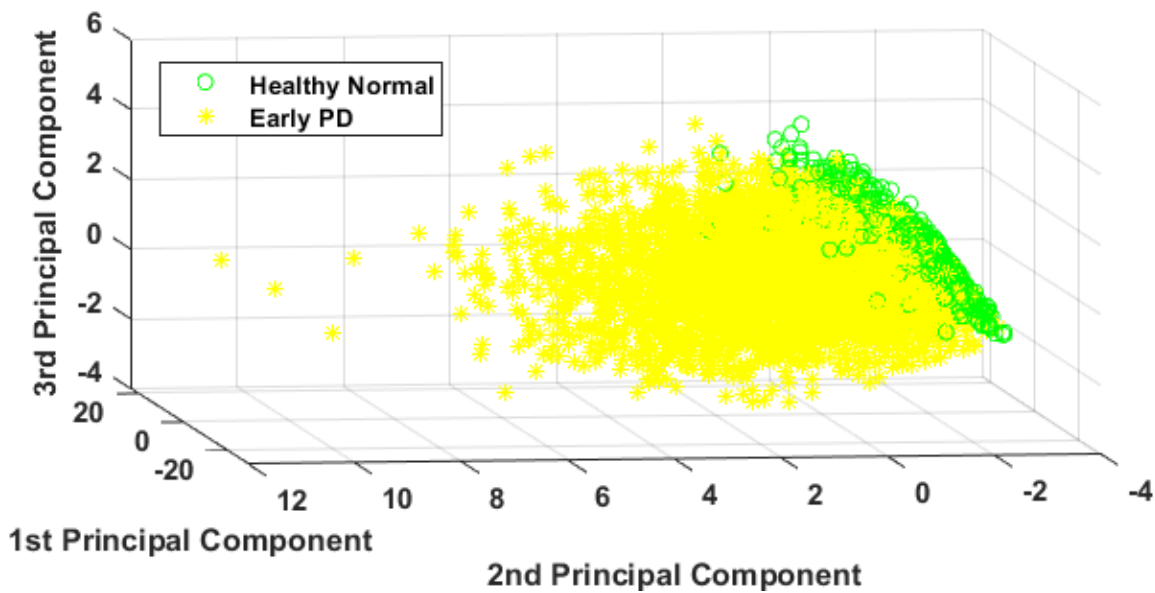

Fig. 5. Visualization of the first three transformed features using PCA

### 3.2 Hyperparameter selection

Hyper parameter optimization for the methods of Boosted Trees, Random Forest and SVM were carried out using Bayesian optimization [54] in a nested cross-validation framework. The Bayesian optimization algorithm attempts to minimize a scalar objective function $f(x)$ for $x$ in a bounded domain. For Boosted trees and SVM, the optimization function was 10-fold cross validation error and for Random forest, the function was the out-of-bag error.

### 3.3 Performance measures of the classifiers used

The performance metrics obtained for the classifiers used is shown in Table 3 below. It is observed that all the classifiers performed with high accuracy and high AUC. To statistically compare the performances from the classifiers, *multiple comparison test*s were carried out [26, 55]. In this procedure, one-way ANOVA followed by post hoc tests using Tukey-Kramer method were carried out for multiple comparison analysis to determine which pairs of means are significant and which are not. The figure below (Fig. 6) shows the plots obtained from these tests. In each plot, the mean of the classifier performance measure is represented by a *circle* symbol and an interval (95% confidence interval) around the symbol. Two means are significantly different if their intervals are disjoint, and are not significantly different if their intervals overlap. In the plots below, solid blue line represent the best performing methods and solid red line represent the groups whose means were significantly different from the best performing method. From these plots, we observe that SVMs were always gave comparable performance with the best performing methods.

**Table 3. Performance measures of the classifiers for the subject-wise and record-wise cases**

(A) Subject-wise nested cross validation

| | Wilcoxon rank sum test | | | | | | | |
|---|---|---|---|---|---|---|---|---|
| | Boosting trees | | Random forests | | Logistic regression | | SVM | |
| | 95% CI | | 95% CI | | 95% CI | | 95% CI | |
| Accuracy | 93.88% | 96.59% | 94.18% | 96.57% | 93.90% | 96.49% | 94.30% | 96.61% |
| Sensitivity | 95.29% | 97.61% | 96.56% | 98.27% | 96.26% | 98.72% | 95.95% | 97.99% |
| Specificity | 83.54% | 95.95% | 79.22% | 92.63% | 81.50% | 89.22% | 82.63% | 94.41% |
| AUC | 96.77% | 98.88% | 96.42% | 98.76% | 96.94% | 98.67% | 96.06% | 98.55% |
| | Feature selection via LASSO | | | | | | | |
| | Boosting trees | | Random forests | | Logistic regression | | SVM | |
| Accuracy | 94.19% | 96.79% | 94.60% | 97.22% | 94.07% | 96.53% | 94.17% | 96.55% |
| Sensitivity | 95.38% | 97.66% | 96.83% | 98.64% | 96.39% | 98.78% | 95.81% | 97.95% |
| Specificity | 85.11% | 96.63% | 80.71% | 94.36% | 81.82% | 89.21% | 82.40% | 94.45% |
| AUC | 96.71% | 98.80% | 96.10% | 98.65% | 96.91% | 98.71% | 96.11% | 98.62% |
| | Feature selection via PCA | | | | | | | |
| | Boosting trees | | Random forests | | Logistic regression | | SVM | |
| Accuracy | 93.81% | 96.39% | 92.99% | 95.64% | 93.97% | 96.47% | 94.23% | 96.77% |
| Sensitivity | 96.45% | 98.36% | 96.77% | 99.09% | 96.38% | 98.82% | 95.86% | 98.14% |
| Specificity | 77.55% | 91.21% | 70.02% | 84.66% | 81.50% | 88.79% | 82.55% | 94.67% |
| AUC | 96.24% | 98.39% | 95.51% | 97.82% | 97.00% | 98.65% | 96.17% | 98.75% |

(B) Record-wise nested cross validation

| | Wilcoxon rank sum test | | | | | | | |
|---|---|---|---|---|---|---|---|---|
| | Boosting trees | | Random forests | | Logistic regression | | SVM | |
| | 95% CI | | 95% CI | | 95% CI | | 95% CI | |
| Accuracy | 94.62% | 96.09% | 95.56% | 96.66% | 94.55% | 96.33% | 95.64% | 96.96% |
| Sensitivity | 95.76% | 97.52% | 97.04% | 98.41% | 97.24% | 98.05% | 96.56% | 98.17% |
| Specificity | 87.40% | 91.23% | 86.17% | 90.87% | 82.18% | 89.58% | 89.04% | 93.58% |
| AUC | 97.61% | 98.48% | 98.04% | 98.77% | 97.53% | 98.44% | 98.09% | 98.82% |
| | Feature selection via LASSO | | | | | | | |
| | Boosting trees | | Random forests | | Logistic regression | | SVM | |
| Accuracy | 95.05% | 96.40% | 96.20% | 97.14% | 94.72% | 96.16% | 95.71% | 96.96% |
| Sensitivity | 96.07% | 97.63% | 97.55% | 98.58% | 97.29% | 98.12% | 96.61% | 98.15% |
| Specificity | 88.90% | 91.94% | 88.31% | 91.92% | 82.50% | 88.73% | 89.27% | 93.56% |
| AUC | 97.72% | 98.59% | 98.59% | 99.29% | 97.63% | 98.51% | 98.05% | 98.86% |
| | Feature selection via PCA | | | | | | | |
| | Boosting trees | | Random forests | | Logistic regression | | SVM | |
| Accuracy | 94.83% | 96.05% | 94.83% | 96.19% | 94.68% | 96.20% | 95.57% | 96.92% |
| Sensitivity | 96.79% | 98.23% | 97.85% | 99.30% | 97.34% | 98.03% | 96.68% | 98.17% |
| Specificity | 83.56% | 87.89% | 78.15% | 84.12% | 82.27% | 89.21% | 88.54% | 92.88% |
| AUC | 97.44% | 98.45% | 97.83% | 98.80% | 97.56% | 98.44% | 97.87% | 98.77% |

*CI represents confidence interval

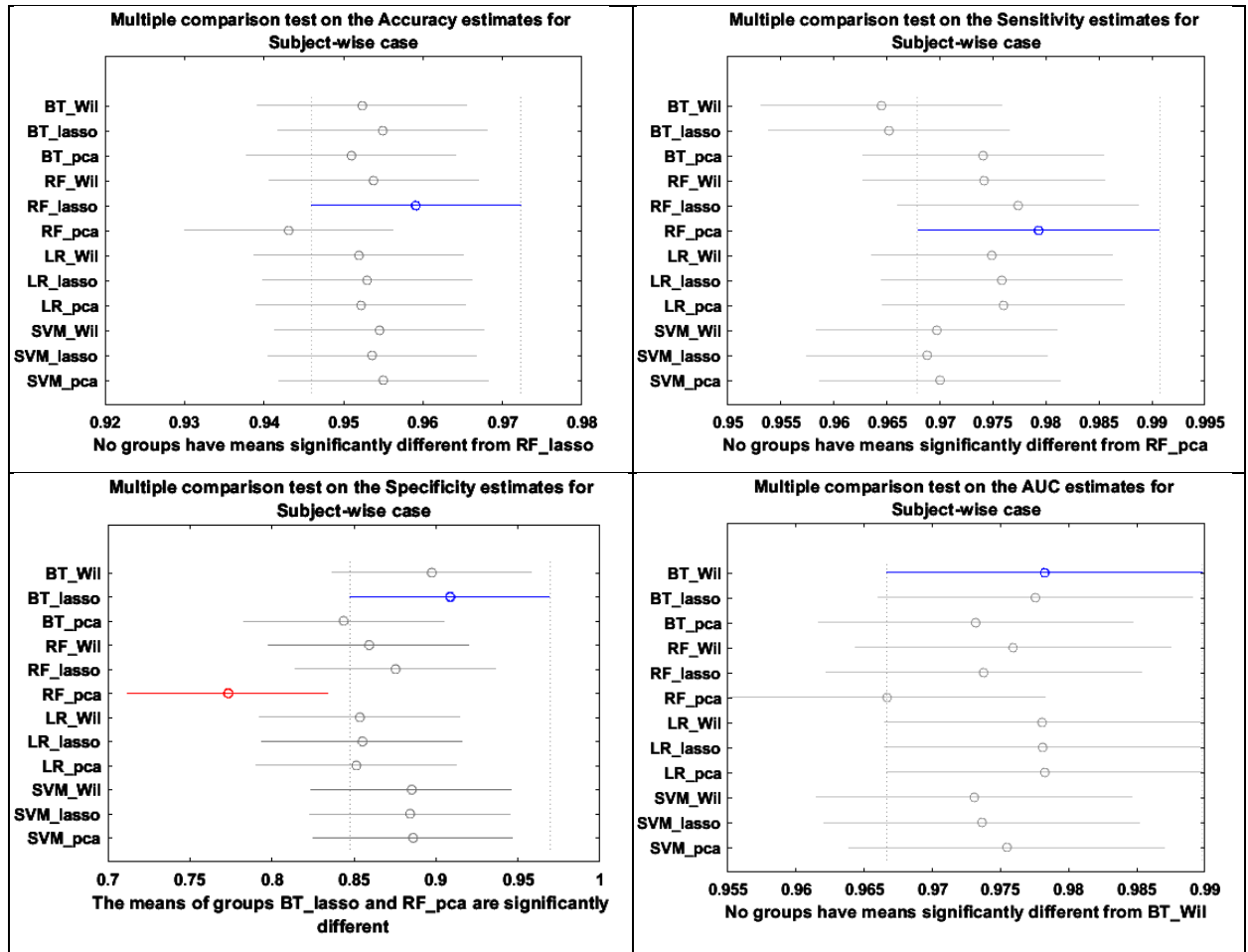

(a) Subject-wise case

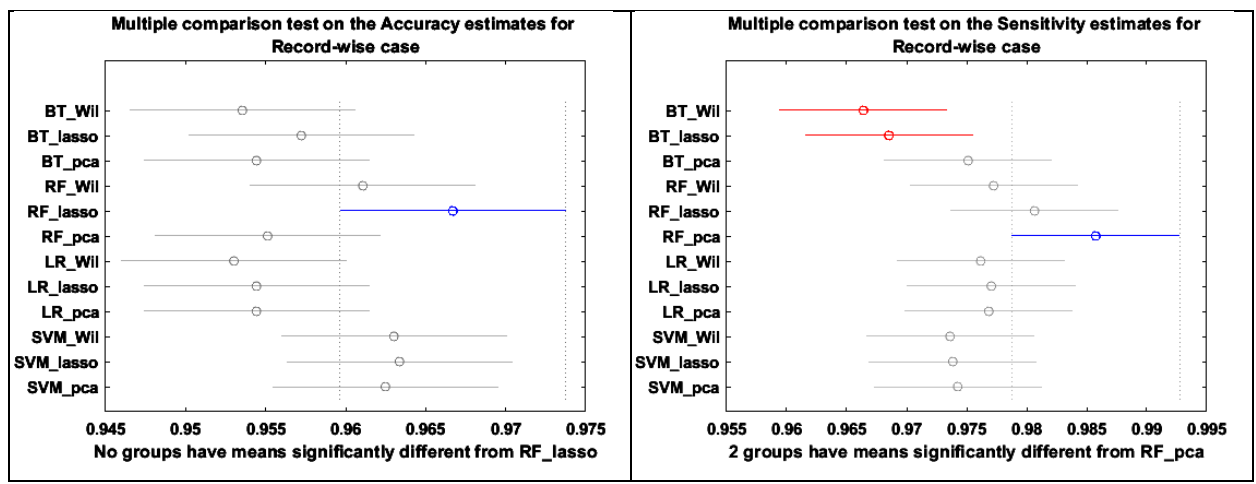

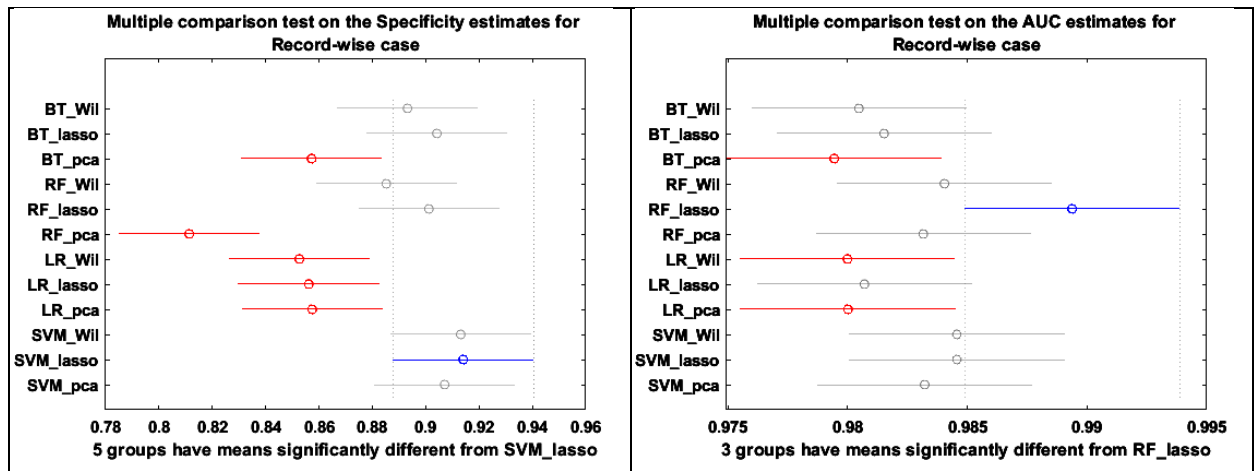

(b) Record-wise case

Fig. 6. Statistical comparison of the classifier performance measures of Accuracy, Sensitivity, Specificity, and Area under ROC (AUC) using multiple comparison tests. BT, RF, LR, and SVM represent Boosted Trees, Random Forests, Logistic Regression, and Support Vector Machine, respectively. Wil, lasso, and pca represent the feature selection techniques used which are Wilcoxon rank sum test, LASSO based, and PCA based feature selection.

**3.4 Comparison of machine learning models with the patient questionnaire scores**

In clinical setting, the total score is an important factor considered. However, there is no threshold that is reported that can be used to distinguish between PD and healthy normal. To compare the performance of questionnaire alone with the machine learning approaches, we have taken the total score of the questionnaire and then plotted the AUCs. The below figure (Fig. 7) shows the AUCs for different approaches, and we can clearly observe that the machine learning approaches are performing much better here.

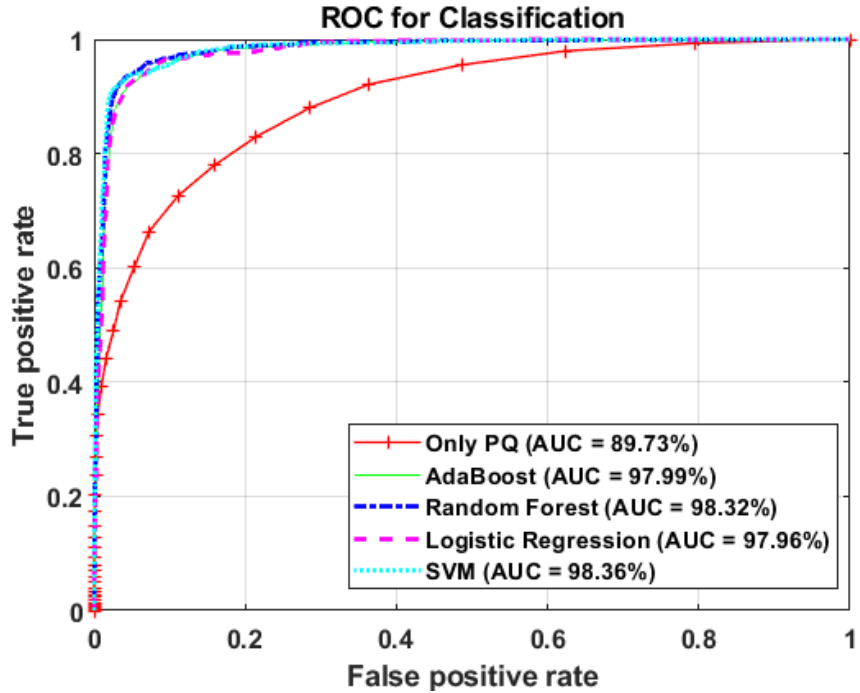

Fig. 7. Plot of Area under ROC curve (AUC) for questionnaire only along with the machine learning approaches.

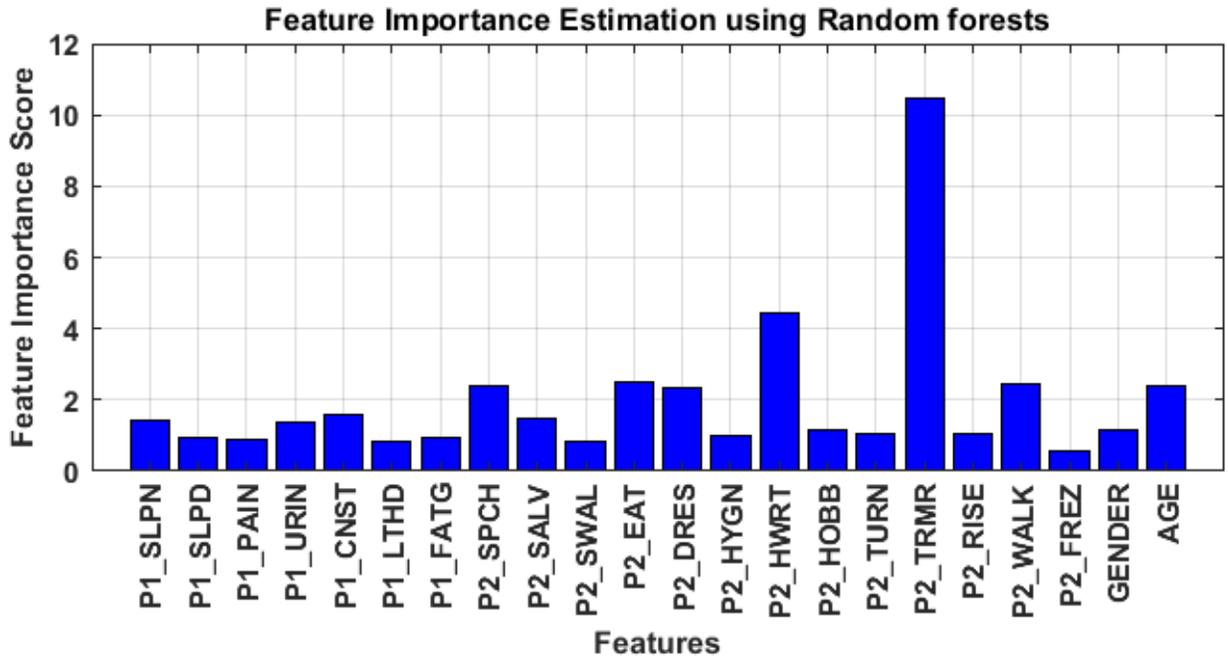

Fig. 8. Plot of feature importance scores from Random forests. It is observed that the features of tremor (P2_TRMR) and handwriting (P2_HWRT) were more important, and freezing (P2_FREZ) was the least

important feature. This pattern is similar to what we observe in Fig 2. It is interesting to note that age was showing higher importance as compared to gender.

### 3.5. Logistic regression model

The performance metrics obtained for logistic regression are shown in Table 3 and it is observed to perform with high accuracy and area under the ROC curve (AUC). The equation $p(y = PD) = (1 + \exp{-(f(x))})^{-1}$ is the logistic model developed where $p(y = PD)$ represents the likelihood of having PD. The logistic model developed using the full data is as given in equation (1) as shown below.

$$\begin{aligned}
f(x) = {} & (4.3677 * Tremor) + (2.2193 * Eating\ tasks) + (2.1455 * Hygiene) \\
& + (1.6894 * Speech) + (1.4171 * Dressing) + (1.1776 * Freezing) \\
& + (1.1211 * Handwriting) + (1.0682 * Constipation\ problems) \\
& + (0.90309 * Chewing\ and\ Swallowing) \\
& + (0.7519 * Saliva\ and\ Drooling) \\
& + (0.72112 * Getting\ out\ of\ bed/car/deep\ chair\ ) \\
& + (0.70782 * Turning\ in\ bed) \\
& + (0.57116 * Doing\ Hobbies\ and\ other\ activities) \\
& + (0.3455 * Walking\ and\ balance) \\
& + (0.16622 * Light\ Headedness\ on\ standing) \\
& + (0.026638 * Daytime\ Sleepiness) + (0.022716 * Urinary\ problems) \\
& - (0.031956 * Age) - (0.33983 * Pain\ and\ other\ sensations) \\
& - (0.41561 * Gender) - (0.41803 * Sleep\ Problems) \\
& - (0.49868 * Fatigue) + 0.54813
\end{aligned} \quad (1)$$

As per this model, the logarithm of odds of an observation to be early PD is positively related to 17 predictors and negatively related to the 5 predictors. It is to be noted that these 5 predictors are negatively related only in the multivariate model and not in the univariate model. The model chi-square value is 3875.3 with $p$-value $<< 0.05$ indicating that the model fits significantly better than a null model (a model with no predictors). The Cox and Snell $R^2$, and Nagelkerke $R^2$ are obtained as 0.49 and 0.82, respectively, which indicate that the relationship between the outcome and predictors is strong, and that the model is useful in risk prediction. These values indicate that the logistic model was well fit to the data.

## 4. Discussion

### 4.1 Note on final models

The main focus of the study was to develop predictive models using patient questionnaire in classifying early PD from healthy normal. Different machine learning techniques were used and compared. All the classifiers performed well and produced comparable performance. Table 3 shows the performance measures obtained for classifiers for subject-wise and record-wise case with feature selection using Wilcoxon rank sum test, LASSO and PCA.

Among all the techniques used, logistic regression is capable of giving the likelihood of PD in terms of a probability. The logistic model using full data is given in Section 3.2. In similar fashion, the Boosted trees, Random forests and SVM models can be built using the full data. The logistic model showed a statistically significant goodness-of-fit measure as indicated by the $R^2$ value. This indicates a good fit to the data, and hence is useful as a predictive model. These prediction models show the potential to aid in the initial evaluation of a subject, which can be followed with clinical evaluation by an PD expert and/or SPECT imaging that has the potential to enhance and validate the diagnostic decision significantly, and thereby can be used a secondary evaluation measure (as it is very expensive).

Table 4 shows a comparison of our study with other related works in developing machine learning based prediction models for PD detection. The contributions (advantages) of this study can be summarised as follows:

a) We improve the works of [32-34] by providing a different perspective of using PQ to develop prediction models for the early PD vs. normal classification.
b) It does not need any special hardware (like in the speech recording or motion recording works) for data acquisition. Instead, it uses a cheap and simple patient questionnaire that can be used even by primary physicians who are not experts in PD.
c) The sample size used is largest as compared to related works in PD detection [3-13, 21-23].
d) The classification accuracy is comparable with related works in PD detection [3-13, 21-23].

**Table 4: Comparison with related works**

| Study | Type of data used | No. of subjects | Classifier used | Accuracy/ AUC |
|---|---|---|---|---|
| [3] | Speech | 23 PD, 8 N | neural network (NN) | 92.9% |
| [4] | Speech | 23 PD, 8 N | parallel feed-forward NN | 91.20% |
| [5] | Speech | 23 PD, 8 N | fuzzy k-nearest neighbor | 96.07%. |
| [6] | Speech | 33 PD, 10 N | SVM | 98.6% |
| [7] | Speech | 20 PD, 20 N | SVM | 77.50% |
| [8] | Motion video | 7 PD, 7 N | linear discriminant analysis and minimum distance classifier | 95.49% |
| [9] | Force tracking | 30 PD, 30 N | SVM | 85% |
| [10] * | SPECT scan | 41 PS, 39 N | SVM | 95% |
| [11] * | SPECT scan | 100 PS, 108 N | SVM | 96.81% |
| [12] | SPECT scan | 369 PD, 179 N | SVM | 96.14% |
| [13] | Smell identification | 193 PD, 157 N | Logistic regression | 88.4% |
| [21] | Demographic, imaging, genetics and clinical | 263 PD + 123 N + 37 SWEDD** | Adboost | 98% |
| [23] | Motor, non-motor and imaging data | 189 PD + 415 N + 63 SWEDD | Probabilistic neural network | >98% |
| [22] | Genetic, non-motor and demographic data | 367 PD + 165 N | Logistic regression | >92% |
| Our study | Patient questionnaire | 446 PD, 180 N | Logistic Regression, Random forests, Boosted trees, SVM | >95% |

PD and N represent Parkinson's disease and healthy normal, respectively. Unlike other approaches, which use sophisticated hardware for signal acquisition or are expensive or use only partial aspects of PD for evaluation, our approach uses PQ which is simple to understand, easy to administer and complete, cheap, extensively tested and evaluates a broad spectrum of features (both motor and non-motor) of PD. The classifier which produced the highest accuracy is shown in 'Classifier used' column. The 'Accuracy/AUC' columns represents either the accuracy or AUC as few studies reported AUC in place of accuracy.

* studies involve Parkinsonism (PS, which is a group of disorders that present PD-like symptoms and PD is one among them) subjects.

**SWEDD stands for Scans Without Evidence of Dopaminergic Deficit. These are the subjects who show PD symptoms but show normal in SPECT scanning,

### 4.2. A note on misclassified instances

To illustrate the misclassified instances, the figures 9 and 10 as given below show the stacked bar charts for the misclassified instances from the Random forests classifier for the subject-wise and record-wise cases, respectively. These plots show the collective misclassified instances from the 100 runs of 10-fold cross-validations. We can see that the pattern for misclassified healthy normals is way different from the severity pattern that we observe in Figure 1. We can clearly observe from the plots that those healthy normals who showed a severity pattern close to that of early PD were misclassified as early PD. And for the early PD case, as the training data involved few healthy normal instances which showed a severity pattern close to that of early PD, the classifier learned to these instances and misclassified few early PD instances as healthy normal. Similar patterns were observed with other classifiers as well. These misclassifications can be reduced by incorporating other features such as inputs from PD experts and/or neuroimaging can increase the performance to a great extent.

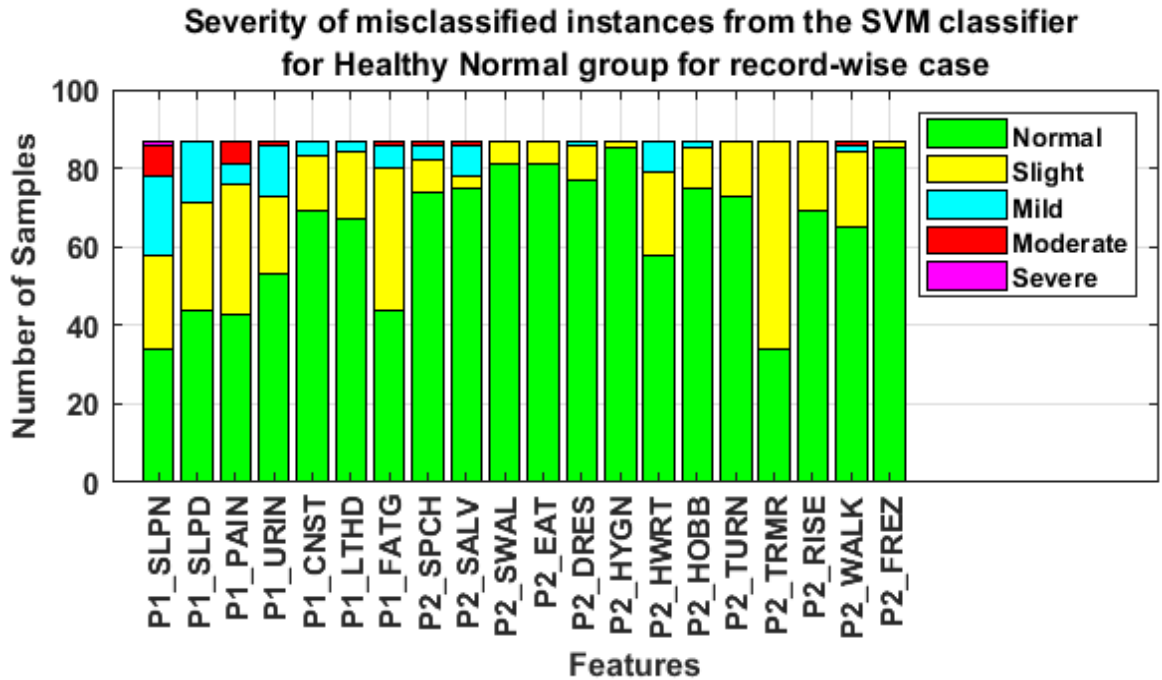

(a)

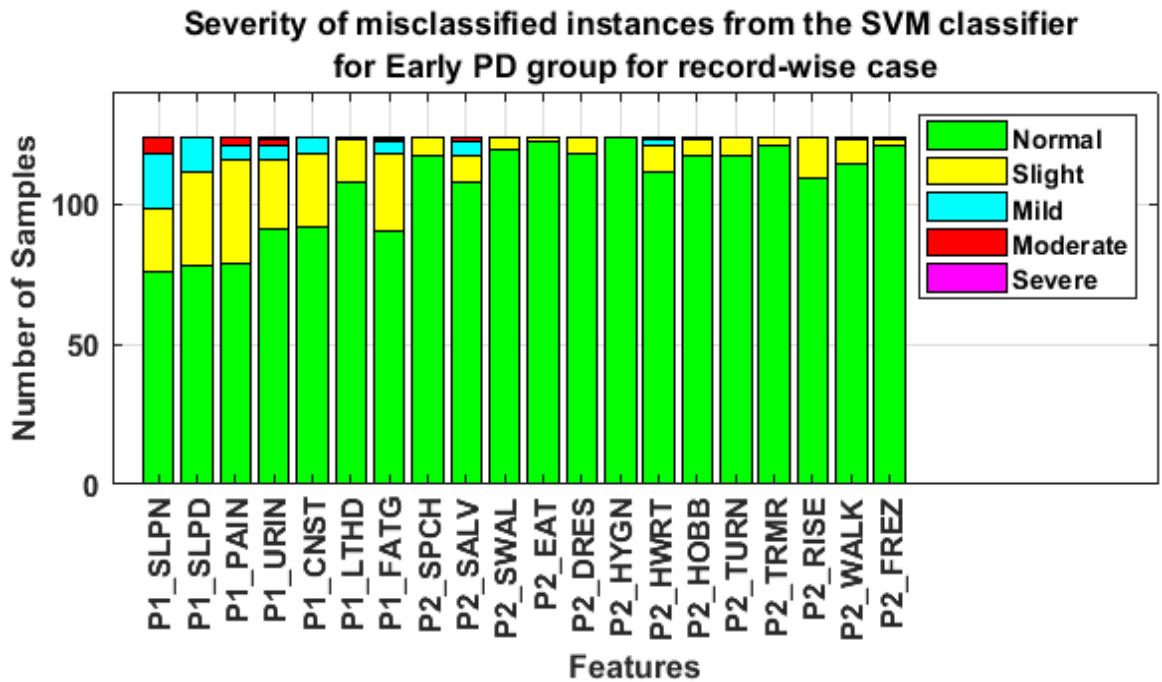

(b)

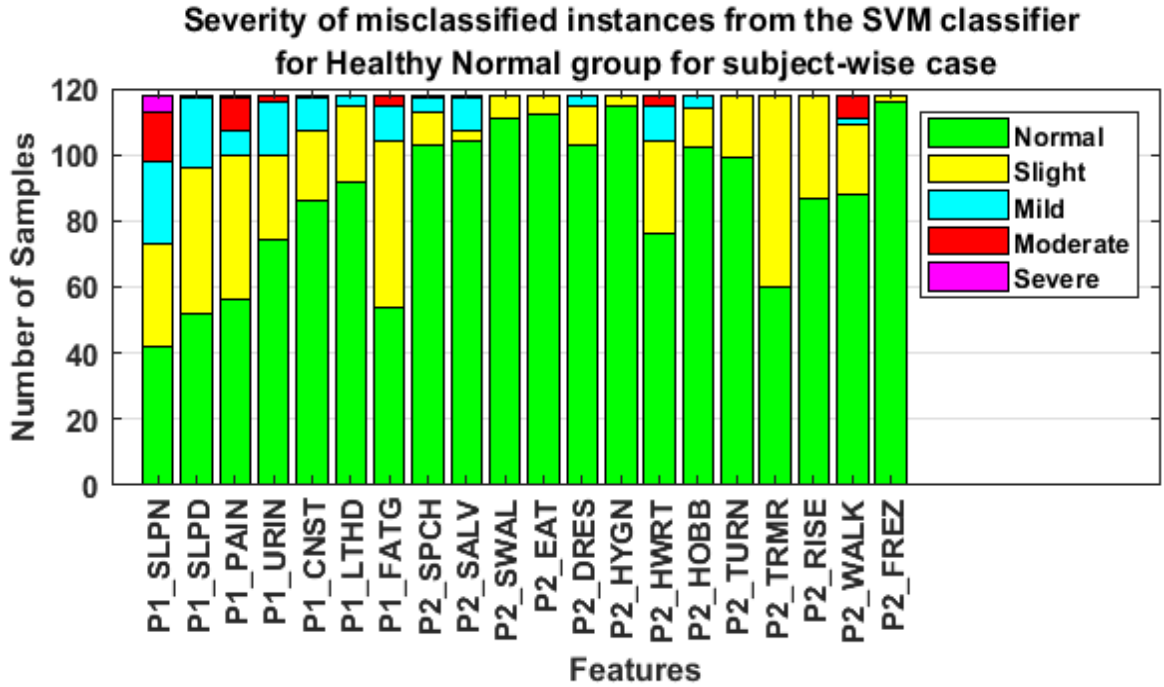

(c)

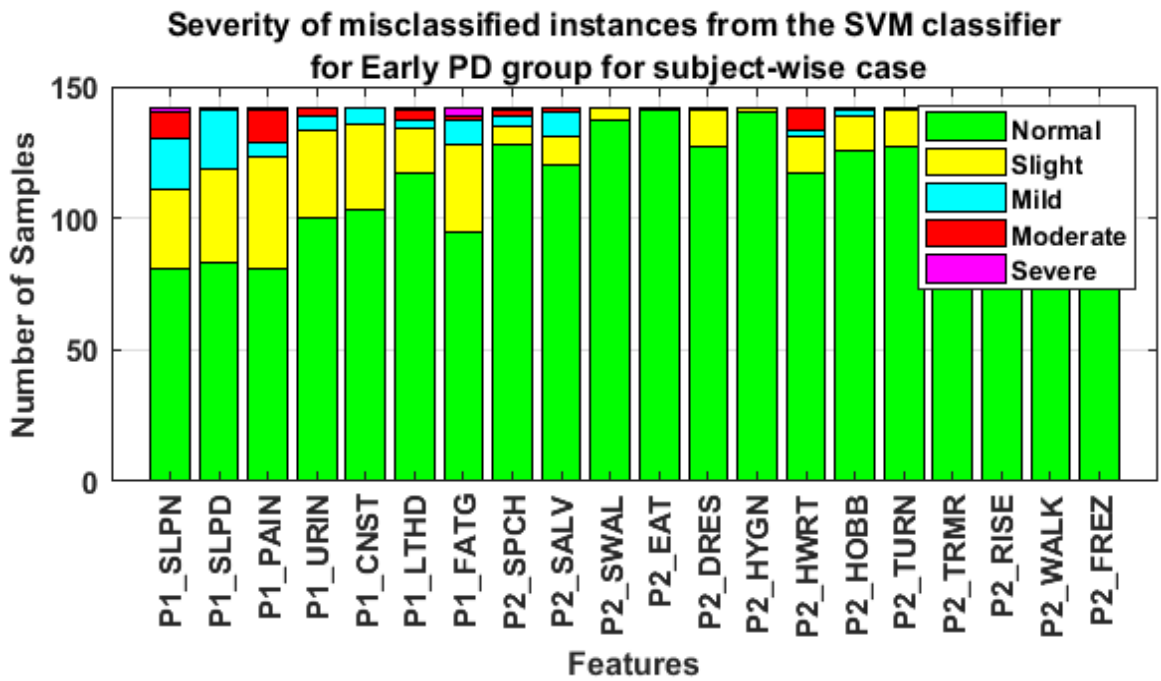

(d)

Fig. 9.The stacked bar charts for misclassified instances from the Random forests classifier for for healthy normal and early PD.

## 4.3. Correlation study of features

Striatal Binding Ratios (SBRs) are clinically used quantitative measurements computed from SPECT imaging for detecting PD. We computed the Spearman correlation coefficient of all features used in the study along with the SBR features with the Hoehn and Yahr (HY) stage of the subjects. The plot of the correlation coefficients is shown in Figure 10 below. It is very encouraging to observe that the 20 features from the PQ used in the study showed significant correlations, although lesser than the SBR values, indicating its usefulness in PD detection. All correlation values were statistically significant, except for the gender feature indicating that gender plays not much role in the severity of PD. Dressing, Handwriting, Getting out of bed/car/deep chair and Tremor were the ones showing higher correlation with HY.

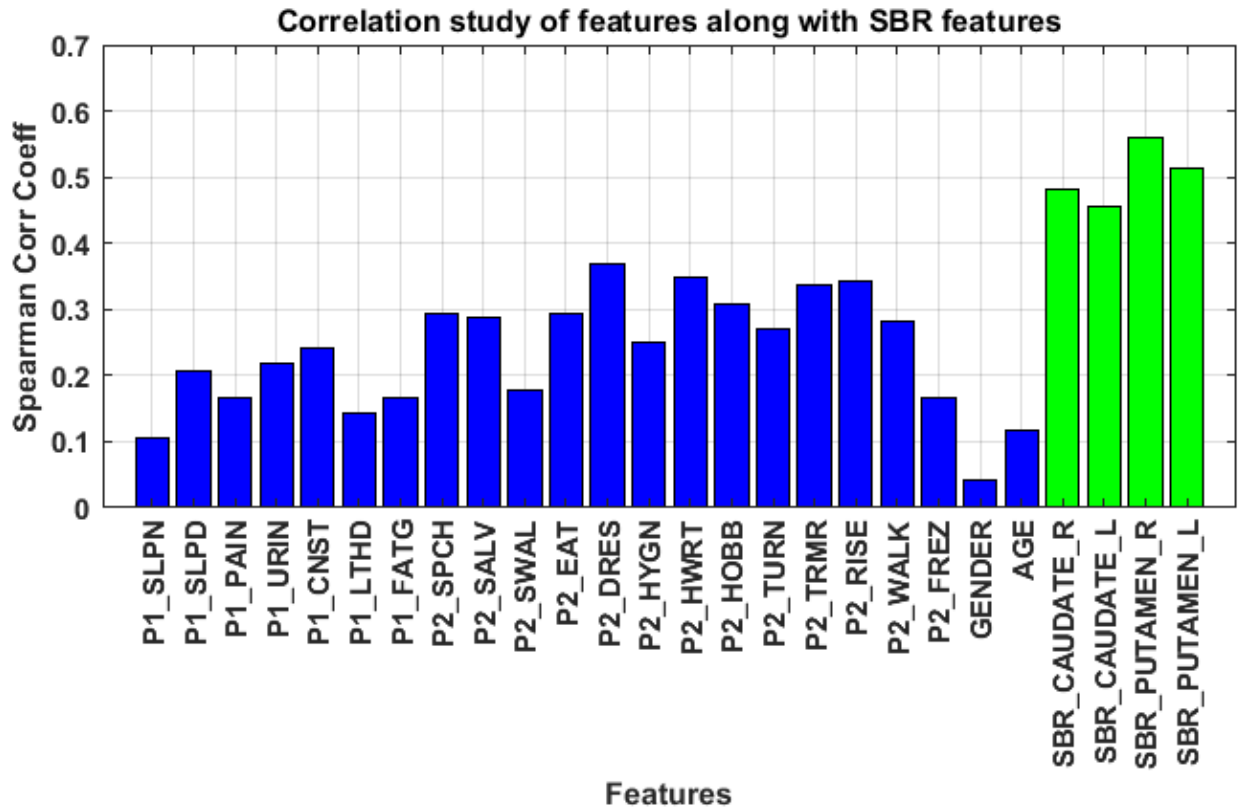

Fig. 10. Correlation study of features along with the clinically used striatal binding ratio (SBR) features. The first 22 features are the ones used in the study (shown in blue). The last 4 features (shown in green) correspond to the SBR ones that are right caudate, left caudate, right putamen and left putamen SBR. The PQ features showed significant correlations although lesser than the clinically used SBR ones, indicating its usefulness in PD detection. Dressing (P2_DRESS), Handwriting (P2_HWRT), Getting out of bed/car/deep chair (P2_RISE) and Tremor (P2_TRMR) were the ones showing higher correlation with HY.

Generally the clinician looks at the total score of the questionnaire for evaluating a subject. The total of the questionnaire features gave a higher correlation of 0.54 with the HY stage.

The correlation of the sum of MDS-UPDRS part III with HY was observed to be 0.76 which is higher as compared to the correlation for the sum of the questionnaire. Incorporation of part III could enhance the performance of the models which is a further extension of the present study.

### 4.4. Note on screening instrument for PD

In the present paper, the prediction models developed could aid a primary physician, who is not a PD expert, to detect possible case of PD and subsequently he/she may refer the subject to an PD expert for confirmatory neurological examination and other expensive tests such as the SPECT scan which provides a specificity as high as 100% [56]. Screening for PD is a clinical need as there exist no approaches as of now that can be used as a screening tool. Developing a screening tool possibly from tool like a patient questionnaire for PD is a long term goal and the presented approaches are like first steps towards that long term goal. The requirements for any screening instrument that can be used for population-wide studies are that it must be sensitive, specific, inexpensive, and non-invasive. An effective screening instrument/tool can serve the following:

i. To identify undiagnosed and preclinical stage PD to initiate early management and therapeutic intervention.

ii. To carry out prevalence studies for evaluating the burden of PD, especially in developing countries where access to health care is limited, and thereby helping policy makers to estimate health care costs and allocate resources more efficiently.

iii. It can aid in clinical trial studies such as the PPMI for recruiting new subjects.

**4.5. Limitations and future work**

A limitation of this study is that the binary classification (early PD/healthy normal) does not provide a differential diagnosis (i.e., diagnosis of PD from other forms of Parkinsonism which show PD-like symptoms such as multiple system atrophy, progressive supra-nuclear palsy and corticobasal degeneration), but it is a promising first step toward that long-term goal. Incorporation of data for these neurodisorders is another possible extension of the present study. Furthermore, sensitivity from a PQ would be much lower as compared to SPECT scans. But our study mainly focused on PQ due to its ease of use and cost-effectiveness. Studies have established that PD patient self-assessment or caregiver evaluation of a patient's disability (based on patient questionnaire) show very close concordance with the neurologist's ratings, and that they are reliable and valid outcome measures [41, 42]. However, a comparison with neurologist's rating, which we have not carried out in this study, can further add value to the work. It is to be noted that a study by Martinez-Martin *et al.* [41] observed that a doctor's rating from a question-and-answer session may bias patient's answers, or underestimate its severity. A self-administered patient questionnaire does not suffer from these biases, and as such, may be a better tool to collect such data.

This proposed approach is an alternative way of joining the items of an existing questionnaire through machine learning. However, it is to be noted that the PPMI group in their research article had acknowledged that the study recruitment is a major challenge for the PPMI due to the difficulty in identifying early-untreated PD subjects. The group, working closely with the Michael J Fox Foundation, had developed a template which they use for recruiting new subjects. The patients have been clinically diagnosed as PD and they form the ground truth or target variable for our study, which enables us to develop predictive models using it. And these models could potentially be used as a clinical aid as well for identifying subjects for clinical trials. Application of this approach for detecting prodromal phase is interesting which is a worthy candidate for future work.

Furthermore, deep learning is an area of machine learning that is increasingly being used in biomedicine. It involves representation-learning methods composing of multiple processing layers to learn representations of data with multiple levels of abstraction. These methods have dramatically improved the state-of-the-art in many supervised classification tasks [57]. These techniques could be applied for generating more complex insights, but for that a substantial amount of data is necessary. Along with this, including sensor data, when it becomes available in the PPMI database, to the problem could further add value to the work.

### 4.5.1. A note on the lack of correlations with subjective-objective data

The items in the questionnaire data which is of subjective type had a weak to moderate correlations (0.10 – 0.38) with the Hoehn and Yahr (HY) stage as we can observe from Figure 11. Although, these correlations were lower, the correlation for the sum of these 20 items came around 0.54 which is a reasonably stronger correlation as compared to individual features.

The SBR data which is of objective type had moderate correlations (0.45 – 0.56) with the HY stage. The total SBRs also had moderate correlations, but slightly higher, with the HY stage. A possible reason for the lower correlations could be that we had used only 3 categories for the HY which are normal, stage 0 and stage 1, rather than using the full spectrum of HY. Another reason could that the progression of the disease is a non-linear process and varies from person to person. However, a further study on the lack of these correlations is a promising future work.

### 4.5.2 Few other recommendations

Literature survey shows that there have been many studies using a variety of data such as imaging data, smell data, movement data, speech data, etc, but only few works focused on developing predictive models using patient questionnaire data. From our study, we demonstrate that effective machine learning models can be developed using the PQ data. In our analysis, we had used the PQ portion of the MDS-UPDRS. This work could be extended by including other patient questionnaires (or any other data) for Parkinson's disease which might improve the performance of the models. Along with that neural network techniques could be used which are capable of learning refined representations from the data, and could possibly perform dimensionality reduction along with performance improvement.

## 5. Conclusion

Computer-aided detection or classification has the potential to reduce inevitable fallibilities and inherent diagnostic variabilities in healthcare, provide guidance and speed up decision-making. The seemingly vast and promising opportunities that machine learning will bring to healthcare will also advance precision medicine and disease management at the level of individual patients. Early detection of PD is an important step to understand the causes, develop better treatments and carry out effective early management of the disease. The importance of diagnostic tools stems from the fact that they can aid in the early detection of PD. In this study, we use the patient questionnaire parts of the widely used MDS-UPDRS to classify early PD cases from healthy normal, to develop prediction models using a variety of classifiers. We observe that these predictive models performed with high accuracy and AUC in distinguishing early PD cases from normal. We infer from the study that such diagnostic models might have the potential to be used as an aid in clinical setting by primary physicians when PD expert are not available, for detecting PD and also for identifying subjects for clinical trials.

**Acknowledgements**

PPMI – a public-private partnership – is funded by the Michael J. Fox Foundation for Parkinson's Research and funding partners, including Abbvie, Avid Radiopharmaceuticals, Biogen Idec, Bristol-Myers Squibb, Covance, GE Healthcare, Genentech, GlaxoSmithKline, Eli Lilly and Company, Lundbeck, Merck & Co., Meso Scale Discovery, Pfizer, Piramal, Hoffmann-La Roche, UCB (Union Chimique Belge), Takeda, Servier, Teva, Sanofi Genzyme and Biolegend.

**References**

[1] S. Fahn, Description of Parkinson's disease as a clinical syndrome, Ann N Y Acad Sci, 991 (2003) 1-14.

[2] M.C. de Rijk, C. Tzourio, M.M. Breteler, J.F. Dartigues, L. Amaducci, S. Lopez-Pousa, J.M. Manubens-Bertran, A. Alperovitch, W.A. Rocca, Prevalence of parkinsonism and Parkinson's disease in Europe: the EUROPARKINSON Collaborative Study. European Community

Concerted Action on the Epidemiology of Parkinson's disease, J Neurol Neurosurg Psychiatry, 62 (1997) 10-15.

[3] R. Das, A comparison of multiple classification methods for diagnosis of Parkinson disease, Expert Syst. Appl., 37 (2010) 1568–1572.

[4] F. Åström, R. Koker, A parallel neural network approach to prediction of Parkinson's Disease, Expert Syst. Appl., 38 (2011) 12470–12474.

[5] H.-L. Chen, C.-C. Huang , X.-G. Yu , X. Xu , X. Sun , G. Wangd, W. Su-Jing, An efficient diagnosis system for detection of Parkinson's disease using fuzzy k-nearest neighbor approach, Expert Syst. Appl., 40 (2013) 263–271.

[6] A. Tsanas, M.A. Little, P.E. McSharry, J. Spielman, L.O. Ramig, Novel Speech Signal Processing Algorithms for High-Accuracy Classification of Parkinson's Disease, IEEE Trans Biomed Eng, 59 (2012) 1264--1271.

[7] B.E. Sakar, M.M. Isenkul, C.O. Sakar, A. Sertbas, F. Gurgen, S. Delil, H. Apaydin, O. Kursun, Collection and Analysis of a Parkinson Speech Dataset With Multiple Types of Sound Recordings, IEEE J. Biomedical and Health Informatics, 17 (2013) 828-834.

[8] C.-W. Cho, W.-H. Chao, S.-H. Lin, Y.-Y. Chen, A vision-based analysis system for gait recognition in patients with Parkinson's disease, Expert Syst. Appl., 36 (2009) 7033-7039.

[9] B.R. Brewer, S. Pradhan, G. Carvell, A. Delitto, Feature selection for classification based on fine motor signs of parkinson's disease, Proc. Ann. Int. Conf. IEEE Engineering in Medicine and Biology Society, EMBC, 2009.

[10] A. Rojas, J.M. Górriz, J. Ramírez, I.A. Illán, F.J. Martínez-Murcia, A. Ortiz, M. Gómez Río, M. Moreno-Caballero, Application of Empirical Mode Decomposition (EMD) on DaTSCAN SPECT images to explore Parkinson Disease, Expert Syst. Appl., 40 (2013) 2756-2766.

[11] I.A. Illan, J.M. Gorrz, J. Ramirez, F. Segovia, J.M. Jimenez-Hoyuela, S.J. Ortega Lozano, Automatic assistance to Parkinson's disease diagnosis in DaTSCAN SPECT imaging, Medical Physics, 39 (2012) 5971-5980.

[12] R. Prashanth, S. Dutta Roy, P.K. Mandal, S. Ghosh, Automatic classification and prediction models for early Parkinson's disease diagnosis from SPECT imaging, Expert Syst. Appl., 41 (2014).

[13] L. Silveira-Moriyama, A. Petrie, D.R. Williams, A. Evans, R. Katzenschlager, E.R. Barbosa, A.J. Lees, The use of a color coded probability scale to interpret smell tests in suspected parkinsonism, Mov Disord, 24 (2009) 1144-1153.

[14] N.Y. Hammerla, J.M. Fisher, P. Andras, L. Rochester, R. Walker, T. Plotz, PD disease state assessment in naturalistic environments using deep learning, Proceedings of the Twenty-Ninth AAAI Conference on Artificial Intelligence, AAAI Press, Austin, Texas, 2015, pp. 1742-1748.

[15] S. Cohen, L.R. Bataille, A.K. Martig, Enabling breakthroughs in Parkinson's disease with wearable technologies and big data analytics, Mhealth, 2 (2016) 20.

[16] A.L. Silva de Lima, T. Hahn, N.M. de Vries, E. Cohen, L. Bataille, M.A. Little, H. Baldus, B.R. Bloem, M.J. Faber, Large-Scale Wearable Sensor Deployment in Parkinson's Patients: The Parkinson@Home Study Protocol, JMIR Res Protoc, 5 (2016) e172.

[17] A.L. Silva de Lima, T. Hahn, L.J.W. Evers, N.M. de Vries, E. Cohen, M. Afek, L. Bataille, M. Daeschler, K. Claes, B. Boroojerdi, D. Terricabras, M.A. Little, H. Baldus, B.R. Bloem, M.J. Faber, Feasibility of large-scale deployment of multiple wearable sensors in Parkinson's disease, PLoS One, 12 (2017) e0189161.

[18] A. Wagner, F. Naama, Y.S. R., A Wavelet-Based Approach to Moniotring Parkinson's Disease Symptoms., arXiv:1701.03161 arXiv preprint (2016).

[19] W. Waks, I. Mazeh, C. Admati, M. Afek, Y. Dolan, A. Wagner, Wrist Sensor Fusion Enables Robust Gait Quantification Across Walking Scenarios, arXiv:1711.06974, arXiv preprint (2017).

[20] W. Maetzler, J. Domingos, K. Srulijes, J.J. Ferreira, B.R. Bloem, Quantitative wearable sensors for objective assessment of Parkinson's disease, Mov Disord, 28 (2013) 1628-1637.

[21] I.D. Dinov, B. Heavner, M. Tang, G. Glusman, K. Chard, M. Darcy, R. Madduri, J. Pa, C. Spino, C. Kesselman, I. Foster, E.W. Deutsch, N.D. Price, J.D. Van Horn, J. Ames, K. Clark, L. Hood, B.M. Hampstead, W. Dauer, A.W. Toga, Predictive Big Data Analytics: A Study of Parkinson's Disease Using Large, Complex, Heterogeneous, Incongruent, Multi-Source and Incomplete Observations, PLoS One, 11 (2016) e0157077.

[22] M.A. Nalls, C.Y. McLean, J. Rick, S. Eberly, S.J. Hutten, K. Gwinn, M. Sutherland, M. Martinez, P. Heutink, N.M. Williams, J. Hardy, T. Gasser, A. Brice, T.R. Price, A. Nicolas, M.F. Keller, C. Molony, J.R. Gibbs, A. Chen-Plotkin, E. Suh, C. Letson, M.S. Fiandaca, M. Mapstone, H.J. Federoff, A.J. Noyce, H. Morris, V.M. Van Deerlin, D. Weintraub, C. Zabetian, D.G. Hernandez, S. Lesage, M. Mullins, E.D. Conley, C.A. Northover, M. Frasier, K. Marek, A.G. Day-Williams, D.J. Stone, J.P. Ioannidis, A.B. Singleton, Diagnosis of Parkinson's disease on the basis of clinical and genetic classification: a population-based modelling study, Lancet Neurol, 14 (2015) 1002-1009.

[23] T.J. Hirschauer, H. Adeli, J.A. Buford, Computer-Aided Diagnosis of Parkinson's Disease Using Enhanced Probabilistic Neural Network, J Med Syst, 39 (2015) 179.

[24] S. Emrani, A. McGuirk, W. Xiao, Prognosis and Diagnosis of Parkinson's Disease Using Multi-Task Learning, Proceedings of the 23rd ACM SIGKDD International Conference on Knowledge Discovery and Data Mining, ACM, Halifax, NS, Canada, 2017, pp. 1457-1466.

[25] Y. Wu, X.Y. Guo, Q.Q. Wei, R.W. Ou, W. Song, B. Cao, B. Zhao, H.F. Shang, Non-motor symptoms and quality of life in tremor dominant vs postural instability gait disorder Parkinson's disease patients, Acta Neurol Scand, 133 (2016) 330-337.

[26] R. Prashanth, S. Dutta Roy, P.K. Mandal, S. Ghosh, High-Accuracy Detection of Early Parkinson's Disease through Multimodal Features and Machine Learning, International Journal of Medical Informatics, 90 (2016) 13-21.

[27] N. Dahodwala, A. Siderowf, M. Baumgarten, A. Abrams, J. Karlawish, Screening questionnaires for parkinsonism: a systematic review, Parkinsonism Relat Disord, 18 (2012) 216-224.

[28] G.U. Hoglinger, I. Rissling, A. Metz, V. Ries, A. Heinermann, H. Prinz, S. Spieker, G. Deuschl, E. Baum, W.H. Oertel, Enhancing recognition of early Parkinsonism in the community, Mov Disord, 19 (2004) 505-512.

[29] N. Sarangmath, R. Rattihalli, M. Ragothaman, G. Gopalkrishna, S. Doddaballapur, E.D. Louis, U.B. Muthane, Validity of a modified Parkinson's disease screening questionnaire in India: effects of literacy of participants and medical training of screeners and implications for screening efforts in developing countries, Mov Disord, 20 (2005) 1550-1556.

[30] K.S. Taylor, C.E. Counsell, C.E. Harris, J.C. Gordon, Screening for undiagnosed parkinsonism in people aged 65 years and over in the community, Parkinsonism Relat Disord, 12 (2006) 79-85.

[31] C.G. Goetz, B.C. Tilley, S.R. Shaftman, G.T. Stebbins, S. Fahn, P. Martinez-Martin, W. Poewe, C. Sampaio, M.B. Stern, R. Dodel, B. Dubois, R. Holloway, J. Jankovic, J. Kulisevsky, A.E. Lang, A. Lees, S. Leurgans, P.A. LeWitt, D. Nyenhuis, C.W. Olanow, O. Rascol, A. Schrag, J.A. Teresi, J.J. van Hilten, N. LaPelle, U.R.T.F. Movement Disorder Society, Movement Disorder Society-sponsored revision of the Unified Parkinson's Disease Rating Scale (MDS-UPDRS): scale presentation and clinimetric testing results, Mov Disord, 23 (2008) 2129-2170.

[32] A.E. Lang, S. Eberly, C.G. Goetz, G. Stebbins, D. Oakes, K. Marek, B. Ravina, C.M. Tanner, I. Shoulson, L.-P. investigators, Movement disorder society unified Parkinson disease rating scale experiences in daily living: longitudinal changes and correlation with other assessments, Mov Disord, 28 (2013) 1980-1986.

[33] D.A. Gallagher, C.G. Goetz, G. Stebbins, A.J. Lees, A. Schrag, Validation of the MDS-UPDRS Part I for nonmotor symptoms in Parkinson's disease, Mov Disord, 27 (2012) 79-83.

[34] C. Rodriguez-Blazquez, J.M. Rojo-Abuin, M. Alvarez-Sanchez, T. Arakaki, A. Bergareche-Yarza, A. Chade, N. Garretto, O. Gershanik, M.M. Kurtis, J.C. Martinez-Castrillo, A. Mendoza-Rodriguez, H.P. Moore, M. Rodriguez-Violante, C. Singer, B.C. Tilley, J. Huang, G.T. Stebbins, C.G. Goetz, P. Martinez-Martin, The MDS-UPDRS Part II (motor experiences of daily living)

resulted useful for assessment of disability in Parkinson's disease, Parkinsonism Relat Disord, 19 (2013) 889-893.

[35] M.M. Hoehn, M.D. Yahr, Parkinsonism: onset, progression and mortality, Neurology, 17 (1967) 427-442.

[36] C.G. Goetz, W. Poewe, O. Rascol, C. Sampaio, G.T. Stebbins, C. Counsell, N. Giladi, R.G. Holloway, C.G. Moore, G.K. Wenning, M.D. Yahr, L. Seidl, Movement Disorder Society Task Force on Rating Scales for Parkinson's Disease, Movement Disorder Society Task Force report on the Hoehn and Yahr staging scale: status and recommendations, Mov Disord, 19 (2004) 1020-1028.

[37] B.K. Scanlon, H.L. Katzen, B.E. Levin, C. Singer, S. Papapetropoulos, A formula for the conversion of UPDRS-III scores to Hoehn and Yahr stage, Parkinsonism Relat Disord, 14 (2008) 379–380.

[38] B.K. Scanlon, H.L. Katzen, B.E. Levin, C. Singer, S. Papapetropoulos, A revised formula for the conversion of UPDRS-III scores to Hoehn and Yahr stage, Parkinsonism Relat Disord, 16 (2010) 151–152.

[39] A. Tsanas, M.A. Little, P.E. McSharry, B.K. Scanlon, S. Papapetropoulos, Statistical analysis and mapping of the unified Parkinson's Disease rating scale to Hoehn and Yahr staging, Parkinsonism Relat Disord, 18 (2012 ) 697-699.

[40] S. Fahn, R.L. Elton, UPDRS program members, Unified Parkinson's Disease Rating Scale, in: S. Fahn, C.D. Marsden, M. Goldstein, D.B. Calne (Eds.) Recent developments in Parkinsons disease, Macmillan Healthcare Information, Florham Park, NJ, 1987, pp. 153–163.

[41] P. Martinez-Martin, J. Benito-Leon, F. Alonso, M.J. Catalan, M. Pondal, A. Tobias, I. Zamarbide, Patients', doctors', and caregivers' assessment of disability using the UPDRS-ADL section: are these ratings interchangeable?, Mov Disord, 18 (2003) 985-992.

[42] R.G. Brown, B. MacCarthy, M. Jahanshahi, C.D. Marsden, Accuracy of self-reported disability in patients with parkinsonism, Arch Neurol, 46 (1989) 955-959.

[43] K. Marek, D. Jennings, S. Lasch, A. Siderowf, C. Tanner, T. Simuni, C. Coffey, K. Kieburtz, E. Flagg, S. Chowdhury, W. Poewe, B. Mollenhauer, P.-E. Klinik, T. Sherer, M. Frasier, C. Meunier, A. Rudolph, C. Casaceli, J. Seibyl, S. Mendick, N. Schuff, Y. Zhang, A. Toga, K. Crawford, A. Ansbach, P. De Blasio, M. Piovella, J. Trojanowski, L. Shaw, A. Singleton, K. Hawkins, J. Eberling, D. Brooks, D. Russell, L. Leary, S. Factor, B. Sommerfeld, P. Hogarth, E. Pighetti, K. Williams, D. Standaert, S. Guthrie, R. Hauser, H. Delgado, J. Jankovic, C. Hunter, M. Stern, B. Tran, J. Leverenz, M. Baca, S. Frank, C.-A. Thomas, I. Richard, C. Deeley, L. Rees, F. Sprenger, E. Lang, H. Shill, S. Obradov, H. Fernandez, A. Winters, D. Berg, K. Gauss, D. Galasko, D. Fontaine, Z. Mari, M. Gerstenhaber, D. Brooks, S. Malloy, P. Barone, K. Longo, T. Comery, B. Ravina, I. Grachev, K. Gallagher, M. Collins, K.L. Widnell, S. Ostrowizki, P. Fontoura, T. Ho, J. Luthman, M.v.d. Brug, A.D. Reith, P. Taylor, The Parkinson Progression Marker Initiative (PPMI), Prog Neurobiol, 95 (2011) 629-635.

[44] G. Bassotti, D. Maggio, E. Battaglia, O. Giulietti, F. Spinozzi, G. Reboldi, A.M. Serra, G. Emanuelli, G. Chiarioni, Manometric investigation of anorectal function in early and late stage Parkinson's disease, J Neurol Neurosurg Psychiatry, 68 (2000) 768-770.

[45] F. Wilcoxon, Individual comparisons by ranking methods, Biometrics bulletin, 1 (1945) 80-83.

[46] R. Tibshirani, Regression shrinkage and selection via the lasso, Journal of the Royal Statistical Society. Series B (Methodological), (1996) 267-288.

[47] J. Friedman, T. Hastie, R. Tibshirani, The elements of statistical learning, Springer series in statistics New York2001.

[48] L. Breiman, Random Forests, Machine Learning, 45 (2001) 5-32.

[49] L. Breiman, J.H. Friedman, R.A. Olshen, C.J. Stone, Classification and Regression Trees, Chapman & Hall (Wadsworth, Inc.), New York, 1984.

[50] Y. Freund, R.E. Schapire, A decision-theoretic generalization of on-line learning and an application to boosting, J. Comput. Syst. Sci., 55 (1997) 119--139.

[51] C. Cortes, V. Vapnik, Support-vector networks, Machine Learning, 20 (1995) 273-297.

[52] S. Dreiseitl, L. Ohno-Machado, Logistic regression and artificial neural network classification models: a methodology review, J Biomed Inform, 35 (2002) 352-359.

[53] C.-C. Chang, C.-J. Lin, LIBSVM: A Library for Support Vector Machines, ACM Trans. Intelligent Systems and Technology, 2 (2011) 1--27.

[54] J. Snoek, H. Larochelle, R.P. Adams, Practical Bayesian Optimization of Machine Learning Algorithms, arXiv:1206.2944, (2012).

[55] M.L. McHugh, Multiple comparison analysis testing in ANOVA, Biochemia medica, (2011) 203-209.

[56] D.L. Jennings, J.P. Seibyl, D. Oakes, S. Eberly, J. Murphy, K. Marek, (123I) beta-CIT and single-photon emission computed tomographic imaging vs clinical evaluation in Parkinsonian syndrome: unmasking an early diagnosis, Arch Neurol, 61 (2004) 1224-1229.

[57] Y. LeCun, Y. Bengio, G. Hinton, Deep learning, Nature, 521 (2015) 436-444.